\documentclass[sigconf,natbib=true,nonacm]{acmart}
\setcopyright{none}
\renewcommand\footnotetextcopyrightpermission[1]{}
\AtBeginDocument{%
  }

\usepackage{graphicx}
\usepackage{subcaption}
\usepackage{dblfloatfix} 
\usepackage{booktabs}
\usepackage{svg}

\begin{document}

\title{Efficient Filtered-ANN via Learning-based Query Planning}


\author{Zhuocheng Gan}
\affiliation{%
  \institution{University of Hawaii Manoa}
  \city{Honolulu}
  \country{USA}}
\email{zgan@hawaii.edu}

\author{Yifan Wang}
\affiliation{%
  \institution{University of Hawaii Manoa}
  \city{Honolulu}
  \country{USA}}
\email{yifanw@hawaii.edu}


\begin{abstract}
Filtered ANN search is an increasingly important problem in vector retrieval, yet systems face a difficult trade-off due to the execution order: Pre-filtering (filtering first, then ANN over the passing subset)  requires expensive per-predicate index construction, while post-filtering (ANN first, then filtering candidates) may waste computation and lose recall under low selectivity due to insufficient candidates after filtering. We introduce a learning-based query planning framework that dynamically selects the most effective execution plan for each query, using lightweight predictions derived from dataset and query statistics (e.g., dimensionality, corpus size, distribution features, and predicate statistics). The framework supports diverse filter types, including categorical/keyword and range predicates, and is generic to use any backend ANN index. 
Experiments show that our method achieves up to 4$\times$ acceleration with $\ge$90\% recall comparing to the strong baselines.
\end{abstract}



\maketitle

\section{Introduction}
\label{sec:intro}
Vector similarity search (nearest neighbors) is a core problem in modern retrieval systems. 
In real world scenarios, retrieval tasks often impose additional constraints on specific attributes of the data. This leads to the \emph{filtered} approximate nearest neighbor (filtered ANN) problem, where the system must return the nearest neighbors from those vectors satisfying a predicate. For example, in an e-commerce search scenario, a user may wish to retrieve items that are semantically similar to their favorites while restricting the results within a price range, i.e., only nearest neighbors that falls in such price range will be returned. 

Although being studied for years~\cite{wu_hqann:_2022, gollapudi_filtered-diskann:_2023, li_sieve:_2025, liang_unify:_2025, li2025attributefilteringapproximatenearest}, filtered ANN remains challenging because the predicate fundamentally changes the effective search space. There are two standard execution strategies for filtered ANN queries: (1) \emph{Pre-filtering} applies the predicate to filter the dataset first and performs ANN search only over the resulting subset. This is effective given low-selectivity predicates, i.e., only a small subset passing the filter. But it often requires expensive per-predicate index construction, as the resulting subsets are different across queries, or it has to use brute-force KNN search on the subset to avoid the index building. Either building index or brute-force search is inefficient especially with high-selectivity filter where the resulting subset is large.  (2) In contrast, \emph{post-filtering} searches nearest neighbors using a global ANN index first and filters the candidates afterwards. Post-filtering avoids per-predicate index maintenance, but can waste computation and suffer recall loss when selectivity is low (few items satisfy the filter), unless the system extends the search to multiple iterations (with efficiency sacrificed) that expands the candidate set aggressively.

Furthermore, these trade-offs are varying across datasets and workloads, making it hard to estimate which execution strategy is better case by case.  Specifically, the best execution plan depends on multiple factors, including dataset scale, distribution, dimensionality, predicate type (categorical vs.\ numeric range), predicate selectivity, and so on. As a result, a ``one-size-fits-all'' filtered ANN strategy can be unreliable, like a method that is fast on high-selectivity predicates may fail to maintain recall under low selectivity, while a conservative strategy that always preserves recall may pay unnecessary overhead for many queries.

We propose a \emph{learning-based query planning framework} that dynamically selects between pre-filtering and post-filtering on a per-query basis. For each filtered ANN query, our planner makes lightweight predictions about which execution plan provides a better efficiency--recall trade-off, based on dataset-level statistics and filter properties. 
At query time, a selectivity estimator first predicts the percentage of data satisfying the filter predicate. This estimated selectivity is then input to the core planner, which selects the optimal execution strategy (between pre and post filtering). Finally the query is executed with the selected strategy.  
Implementation details are in Section~\ref{sec:exp-setting}.  

Our proposed framework is generic: (1) it supports both categorical label and numeric range predicates, which covers most use cases; (2) it can be deployed on top of any existing indexes and execution strategies without constraints. And the training/predicting process is lightweight, not requiring GPU. This ensures the framework raises minimum overhead and cost. Furthermore, our evaluation shows its performance is outstanding with significantly high speed given strict recall requirement. To our best knowledge, our work is the first study that builds a learned query planner/optimizer for filtered ANN queries. Our contributions are as follows:
\begin{itemize}
    \item We introduce a novel \emph{learning-based query planner} for filtered ANN search that dynamically selects between pre- and post-filtering on a per-query basis. 
    \item We design lightweight selectivity estimation and predictive planning approaches which are effective and efficient.
    \item We conduct extensive experiments to evaluate the proposed method which show outstanding performance.
\end{itemize}

\section{Related Work}
\label{sec:related}
Prior work of filtered ANN has explored predicate-aware ANN index designs, hybrid query engines, and adaptive execution strategies to address the challenges posed by varying filter selectivity. Graph-based extensions such as ACORN~\cite{patel_acorn:_2024}, Filtered-DiskANN~\cite{gollapudi_filtered-diskann:_2023}, and HQANN~\cite{wu_hqann:_2022} incorporate predicate awareness into proximity graph traversal to reduce wasted computation without constructing per-predicate indexes. More recent systems focus on robustness under range filtering and shifting predicates: UNIFY~\cite{liang_unify:_2025} proposes a unified graph structure supporting multiple filtering strategies, NaviX~\cite{sehgal_navix:_2025} designs a predicate-agnostic vector index for graph database workloads, WoW~\cite{wang_wow:_2025} targets efficient range-filtered search with incremental indexing, and SIEVE~\cite{li_sieve:_2025} selects from a collection of indexes using analytical cost modeling to meet recall targets. In parallel, vector databases and hybrid query engines such as Milvus~\cite{wang_milvus:_2021} and CHASE~\cite{ma_chase:_2025} integrate vector search with structured query processing, highlighting the need for optimization and planning beyond the ANN index itself. In contrast to prior work that primarily focuses on improving the index structure design, our approach introduces an external, lightweight, learning-based planner that dynamically chooses the optimal execution strategy with given index. It supports any types of ANN indexes and execution strategies. And our method is easy to maintain and robust since it does not touch the complex internal of index structures. 

\section{Learning-based Query Planning}
\label{sec:planner}

In this paper, a filtered ANN query can be represented as a triple $(Q, P, k)$, where $Q$ is a query vector, $P$ is a predicate over vector metadata, and $k$ is the required number of returned results. Each vector is associated with a record of multiple attributes in the metadata, where some attributes are categorical like color while some are numeric like age. A predicate is a condition over specific attributes, like ``age > 20''. Such a filtered ANN query will return the top-$k$ closest vectors to $Q$ which satisfies $P$.     

\begin{figure}[h]
    \vspace{-15pt}
    \centering
    \includegraphics[width=\linewidth]{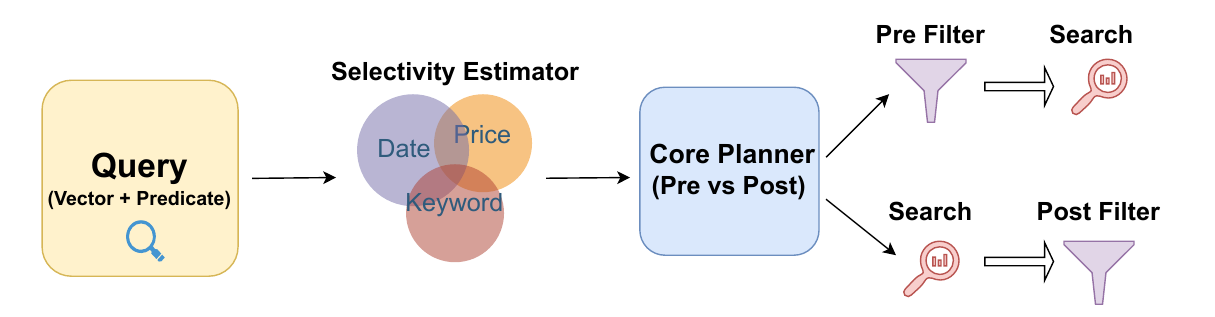}
    \caption{Workflow of the learned query planning framework}
    \label{fig:workflow}
    \vspace{-15pt}
\end{figure}

Filtered ANN queries can be evaluated using different execution strategies, most notably pre-filtering and post-filtering. Our goal is to dynamically select between pre-filtering and post-filtering execution strategies for filtered ANN queries. As shown in Figure~\ref{fig:workflow}, the proposed framework consists of two main components: (1) \emph{selectivity estimator} that accepts a filter/predicate with dataset statistics and estimates its selectivity, and (2) \emph{core planner} that accepts a filtered ANN query with dataset statistics and determines the optimal execution strategy.

\subsection{Training Data Preparation}
\label{sec:threshold}

To train the core planner, on each dataset, we construct a set of filtered ANN queries with controlled selectivity, with the fraction of vectors satisfying the filter ranging from 1\% to 25\%. For each query, we evaluate both execution strategies and compute a utility score:
$U = \frac{\mathrm{Recall@}k}{T_{\text{search}}}$,
where $T_{\text{search}}$ denotes the total end-to-end query processing time (including filtering, ANN search, and any candidate expansion). We use $U$ to indicate which execution strategy is better: a higher $U$ stands for a better performance considering both quality and speed. 
By such, we identify which execution strategy is better for each training query above, acting as the groundtruth/label for the query. The training queries and the labels are then used to train the core planner: given a input query, the core planner chooses the better strategy.  

To train the selectivity estimator, we use the filters from the same training queries above, and the groundtruth is the selectivity of each filter. 

In addition, each training query comes with the necessary dataset statistics, like dataset scale, dimension, histograms of attribute values, etc. 

\subsection{Selectivity Estimation}
\label{sec:selectivity}
Accurate selectivity estimation is critical for execution planning. We handle categorical (keyword) predicates and numeric range predicates separately, and also model their interactions. In this section, the terms "label" and "keyword" both refer to the value of a category attribute, like "green" for attribute "color".  

\subsubsection{Categorical (Keyword) Predicates}
\label{sec:selectivity-keyword}

For single-label(keyword) predicates, like ``color = green'', selectivity is estimated using precomputed frequency statistics, i.e., dictionary of each category's frequency in the dataset. The frequency of a label/category, like "green" in "color" attribute, is calculated as the number of its occurrences over the total number of data points.  
Such frequencies are the ground truth selectivities for single-label predicates, and the selectivity estimation for single-label is completed by looking up the dictionary. 

For conjunctions of two labels, like "color = green AND type = shoes", we use a precomputed 2D co-occurrence matrix, where each entry $(i,j)$ represents the fraction of data points containing both label $i$ and label $j$, i.e., the joint frequency for each two labels. The selectivity estimation is completed by looking up this matrix. 

For conjunctions involving three or more labels, the exact estimation like above becomes expensive and pairwise statistics are insufficient to capture complex interactions. 
To address this, we design a lightweight random sampling approach and train a Gradient Boosting Model to refine selectivity estimates beyond two labels. 
First, we sample a subset of data points from the dataset. All selectivities/frequencies below will be calculated within this subset. In practice, we find that sampling 1–5\% of the dataset is sufficient to capture multi-label interactions with minimal overhead.
Then the model is trained with 300 estimators, maximum depth 4, and a learning rate of 0.05. The model takes as input a set of lightweight features, including (1) the selectivity assuming all labels are independent, (2) joint label selectivity for each two queried labels, and (3) average Pointwise Mutual Information (PMI) for all two-label pairs among the queried labels, to capture the pairwise label dependencies. PMI is defined as 
$\mathrm{PMI}(x,y)=\log\frac{P(x,y)}{P(x)P(y)}$, where x and y are two labels. The gradient boosting regressor learns a nonlinear correction function over these features to approximate the true multi-label selectivity.
Particularly, the model's input also includes additional range-predicate-related features, which is introduced in Section~\ref{sec:selectivity-mix}. But we short circuit them (e.g., setting queried range to be zero) for label-only predicates.   

\subsubsection{Numeric Range Predicates}
\label{sec:selectivity-range}

For numeric range predicates, we estimate selectivity using precomputed histograms on the numeric attributes. Selectivity is approximated by summing the histogram bins covered by the queried range, with partial contributions for boundary bins. For boundary bins that are only partially overlapped with the query range, we assume a uniform distribution within the bin and multiply the bin count proportionally to the fraction of the bin width covered by the range. 
For multi-range predicates, we only consider the cases that all ranges are over the same attribute, like "(age > 20 AND age < 25) OR age < 10". 
In such cases, the full query range is the union of all non-continuous ranges. In the example above, the non-continuous ranges are (20, 25) and [0, 10). Given the full query range, the selectivity is estimated in the same way, i.e., summing the covering histogram bins. 
In our evaluation, we find that using 1,024 histogram bins accurately captures the distribution of range predicates. We will expand the proposed selectivity estimation method to more complex range predicates in future work. Note that for pure range predicates, we do not use the trained model above, as histograms are enough. The trained estimator is used in categorical as well as the mixed predicates (introduced below).

\subsubsection{Mixed Predicates}
\label{sec:selectivity-mix}
For mixed filters that include both categorical and range predicates, the interactions between range and label are captured using label-range pairwise statistics (i.e., label-range co-occurrence matrix, similar to the two-label case). The label-range pairwise joint selectivity is then computed from the matrix. Finally the aforementioned selectivity estimator model (Section~\ref{sec:selectivity-keyword}) is called, with input features including (1) range-predicate-related features: histogram-based range predicate selectivity, total width and midpoint of the range spans, and sum of the label-range pairwise joint selectivity, as well as (2) the label-predicate-related features mentioned in Section~\ref{sec:selectivity-keyword}. 


\subsection{Core Planner: Choosing Execution Strategy}
\label{sec:decision}
Given estimated selectivity and dataset characteristics, the core planner predicts the optimal execution strategy. We train a two-layer MLP classifier that makes an execution strategy decision (pre-filter or post-filter) based on query-related and dataset-related features, including dataset statistics (dimensionality, corpus size, and vector distribution measure), and estimated filter selectivity from the query. The model output is a binary decision indicating whether pre-filtering or post-filtering is expected to yield higher utility. 

The MLP consists of two hidden layers of width 64 and 32, with ReLU activation and Softmax. It is trained for up to 500 epochs with batch size 200, using Adam (learning rate $10^{-3}$) with $L_2$ regularization and early stopping. Hyperparameters are selected via a small grid search using cross-validation and ROC-AUC as the objective. The model is trained independently per dataset and incurs minimal inference overhead at query time. 


\section{Evaluations}
\subsection{Experiment Settings}
\label{sec:exp-setting}
All experiments are conducted on a machine equipped with an Intel Core Ultra 9 CPU and 32\,GB of RAM via WSL2 Ubuntu system. 
All methods are evaluated on single CPU, and no GPU is used.

\noindent\textbf{Datasets}: 
We evaluate our learned query planning method through a series of experiments on both real and synthetic datasets. The statistics of datasets are shown in Table~\ref{tab:datasets} and as follows:

    \textbf{Real-World Datasets}: We use two real-world filtered KNN datasets that include both vectors, filters and filtered KNN queries. 
    (1) \textbf{ArXiv}~\cite{qdrant_qdrant/ann-filtering-benchmark-datasets_2026} contains 2.14M Arxiv paper embeddings (384-dim). The dataset includes real metadata and both numeric range and keyword-based filtered ANN queries.
    (2) \textbf{Wolt}~\cite{qdrant_qdrant/wolt-food-clip-vit-b-32-embeddings_nodate} is a food image dataset containing 1.72M image embeddings of 512 dimensions, as well as metadata about each food image. 
    It does not provide filtered ANN queries, so we select some attributes of the metadata to construct the queries and groundtruth. 
    
     \textbf{Synthetic Datasets}: 
    We also construct two filtered KNN datasets based on pure KNN datasets without native metadata. 
    (1) \textbf{GloVe-200} is a filtered KNN dataset built on top of the GloVe 200-dim dataset~\cite{bernhardsson_erikbern/ann-benchmarks_2026} (a text embedding dataset including 1.18M 200-dim word embeddings). As the original dataset does not include native metadata (attributes) and filtered queries, we generate numeric attributes for each embedding, based on which we further compose numeric-range filtered KNN queries with selectivities varying from 1\% to 20\%.
    (2) \textbf{SIFT} is a filtered KNN dataset built on top of the SIFT-1M dataset~\cite{bernhardsson_erikbern/ann-benchmarks_2026} (an image embedding dataset consisting of 1M 128-dim vectors). The original dataset also lacks native metadata and filtered queries, so we generate attributes and queries the same as GloVe-200.

\begin{table}[t]
\centering
\caption{Dataset Statistics}
\label{tab:datasets}
\begin{tabular}{lrrrl}
\toprule
Dataset & Size & Dimension & Filter type \\
\midrule
ArXiv & 2.14M & 384 & Mixed (real range + keyword) \\
Wolt & 1.72M & 512 & Range (real-valued) \\
GloVe200 & 1.18M & 200 & Range (synthetic) \\
SIFT & 1.00M & 128 & Range (synthetic) \\
\bottomrule
\end{tabular}
\vspace{-12pt}
\end{table}

\begin{figure*}[t]
    \vspace{-10pt}
  \centering
    \begin{subfigure}[t]{0.24\textwidth}
    \includegraphics[width=\linewidth]{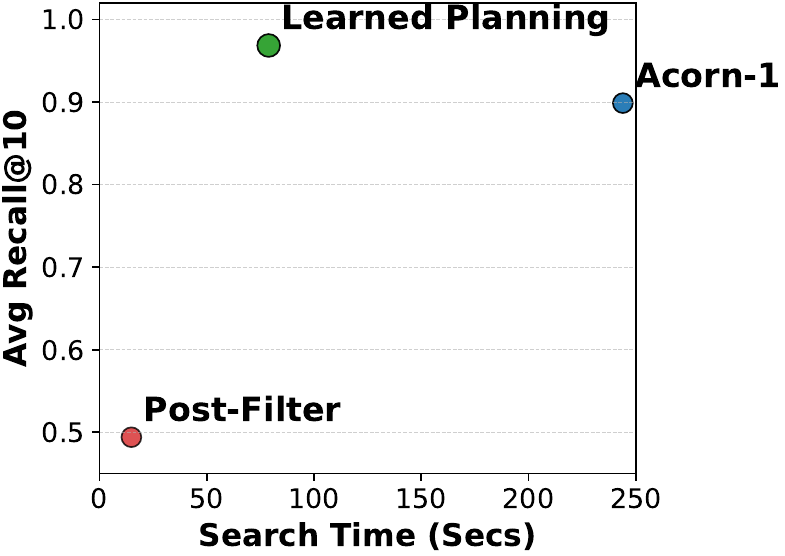}
    \caption{ArXiv 2.13M}
  \end{subfigure}
  \begin{subfigure}[t]{0.23\textwidth}
    \includegraphics[width=\linewidth]{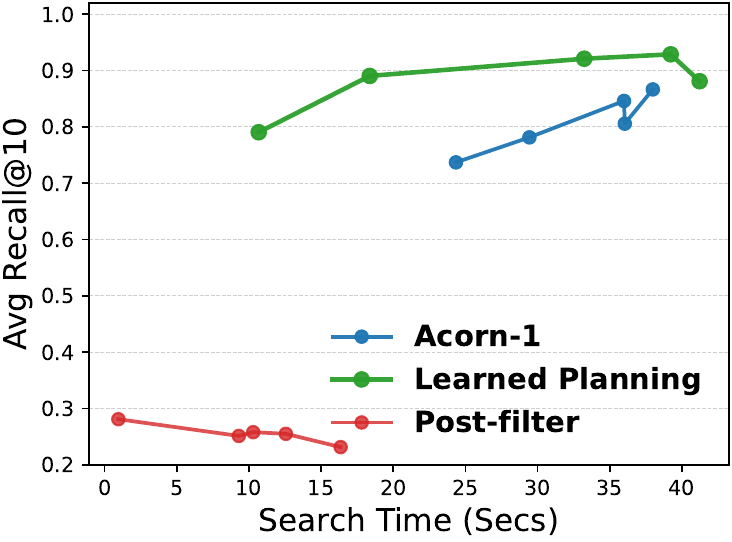}
    \caption{Glove 200 1.18M}
  \end{subfigure}\hfill
  \begin{subfigure}[t]{0.24\textwidth}
    \includegraphics[width=\linewidth]{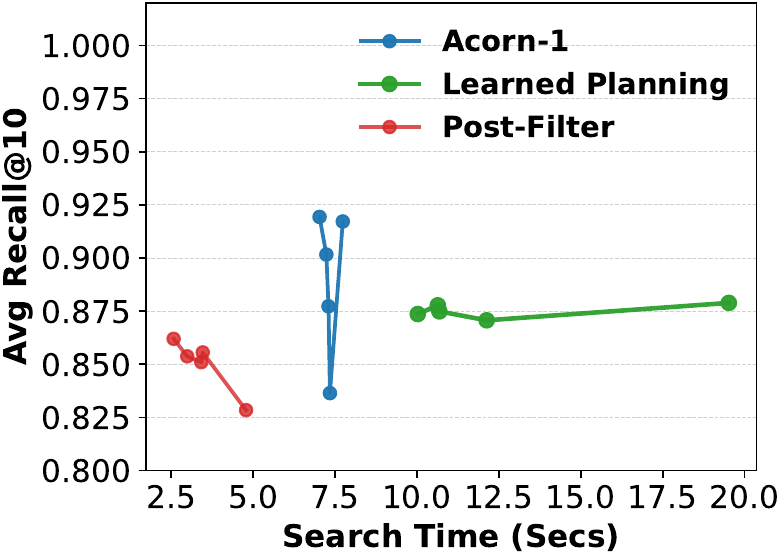}
    \caption{Wolt 1.72M}
  \end{subfigure}\hfill
  \begin{subfigure}[t]{0.24\textwidth}
    \includegraphics[width=\linewidth]{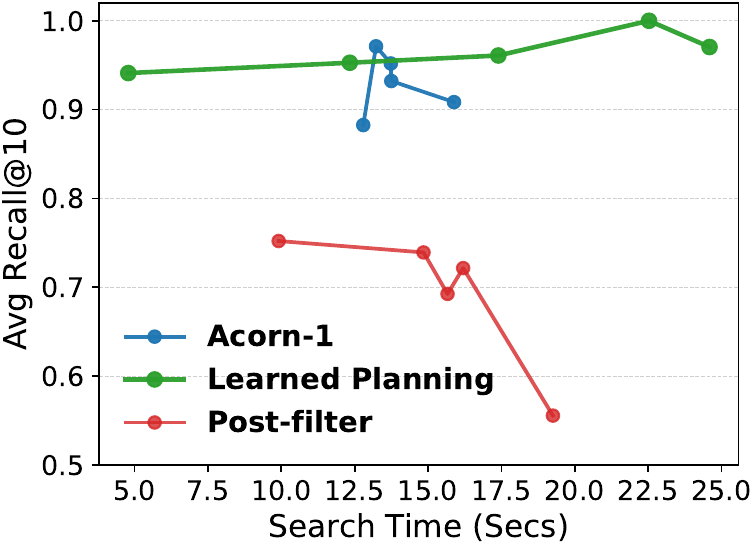}
    \caption{SIFT 1M}
  \end{subfigure}\hfill

  \caption{Latency--recall trade-offs across datasets under varying selectivity.}
  \captionsetup{skip=1pt}
  \label{fig:main-results}
  \vspace{-12pt}
\end{figure*}

\noindent\textbf{Methods}: We evaluate our proposed learned query planning against three baseline methods as follows.

(1) \textbf{Pre-Filtering:}
Pre-filtering applies the filter prior to nearest neighbors search, restricting the search space to vectors that satisfy the filter. After filtering, we perform brute-force search over the resulting subset. This is because high-performance indexes like HNSW usually requires long construction time, causing the evaluation unacceptably time-consuming, especially in high-selectivity circumstance. For example, in our initial evaluation, pre-filtering with HNSW index could take more than 4000 seconds for end-to-end query processing while it only takes 850 seconds with brute-force search (no construction time). Therefore we implement pre-filtering with brute-force KNN in this paper.     

(2) \textbf{Post-Filtering:}
Post-filtering performs ANN search over a global index (built at initialization) and applies the filter to the retrieved candidates. To mitigate recall loss under low selectivity, we expand the candidate set by a factor of $\alpha$ (i.e., retrieving $\alpha k$ candidates). If fewer than $k$ valid results remain after filtering, we iteratively double $\alpha$ and repeat the search and filtering until at least $k$ filtered neighbors are obtained.

(3) \textbf{ACORN-1~\cite{patel_acorn:_2024}:}
ACORN-1 is a strong filtered ANN baseline. ACORN-1 integrates the filtering into a graph-based ANN index. It is neither pre nor post filtering, but filtering during the ANN search. We tune the hyperparameter $M$ over the recommended search grid, and report the best latency--recall results.

(4) \textbf{Learned Query Planning:}
Our learned planner is deployed on top of the pre-filtering and post-filtering baselines described above. For each query, it predicts the appropriate execution strategy in real-time, and executes the query using the selected baseline implementation.   

\noindent\textbf{Metrics: }
The metrics are \textit{recall} (for search quality) and \textit{end-to-end running time} (for efficiency). The running time includes planning overhead (model inference time) and the actual search process, excluding the planner training time and index building time which are completed offline. We report average recall and average time over all queries for each dataset. 

\subsection{Evaluation Results}

\subsubsection{Index Construction Cost}
Comparing to the state-of-the-art methods that usually design a new index structure, our approach has a critical advantage: it is significantly more efficient to build. 
Table~\ref{tab:build_time} reports construction time for ACORN-1 versus our method (including model training and the global index building in post-filtering). Across all datasets, our approach significantly reduces building overhead, achieving a speedup ranging from $1.24\times$ to $20.24\times$. The time saving is more significant on larger datasets (e.g., ArXiv and GloVe-200). 

\subsubsection{End-to-End Results.}

Figure~\ref{fig:main-results} reports latency–recall trade-offs across all datasets. For each of the three datasets except ArXiv, we adjust the predicates in its evaluation queries such that the average filter selectivity varies from 1\% to 20\%, and measure the performance of all methods under each different average selectivity. Since ArXiv provides filtered ANN queries natively, we do not adjust it, and each method only has one performance measure rather than a curve. The results are shown in Figure~\ref{fig:main-results}.      

Pre-filtering achieves 100\% recall but incurs substantially higher latency due to the slow ANN search (without index) or heavy index building on filtered subset, often one to two orders of magnitude slower than other methods. Specifically, the pre-filtering end-to-end time is: 888.81s on ArXiv, 600.44s on Wolt, 182.24s on GloVe-200, and 27.36s on SIFT. So we do not report pre-filtering results in the figure. 
Among the other three methods, post-filtering achieves low latency and low recall, while ACORN-1 often sacrifices speed to achieve high recall, i.e., high latency and high recall. In contrast, our learned planner consistently achieves high recall at substantially lower latency, showing the best performance in most cases.
On ArXiv, learned planner achieves $0.96$ recall@10 with up to 4$\times$ acceleration over ACORN-1 and 10$\times$ speedup over pre-filtering. Similar trends hold on GloVe-200, where our method requires significantly less latency to achieve same recall as ACORN-1, e.g., 30\% time of ACORN to reach the 0.8 recall. On SIFT, our method is overall the best performed, too. 
On Wolt, post-filtering already achieves strong recall at very low latency, indicating that the workload is more favorable to post-filtering. In this case, ACORN-1 improves recall with higher latency, while learned planner achieves competitive recall to post-filtering but not outperforming ACORN-1.  This indicates that when any execution strategy (like post-filtering here) is near-optimal for a workload, learned planner has limited opportunity to further improve the performance. But this does not affect the usefulness of our method: since the real-world queries and datasets are diverse and dynamic, one execution strategy is hard to be always dominant.     

Overall, these results support the need for adaptive execution: the optimal filtered-ANN strategy varies substantially across datasets and filter workloads. The learned planner provides consistent improvements in most cases, while remaining competitive where strong baselines already perform well.

\begin{table}[t]
\centering
\small
\caption{Method construction time (seconds)}
\label{tab:build_time}
\begin{tabular}{lrrr}
\toprule
\textbf{Dataset} & \textbf{ACORN-1} & \textbf{Learned Planner} & \textbf{Speedup} \\
\midrule
Wolt      & 492.37  & 397.95  & 1.24$\times$ \\
SIFT      & 146.29  & 35.37   & 4.14$\times$ \\
GloVe-200  & 917.44  & 73.22   & 12.53$\times$ \\
ArXiv     & 9739.47 & 481.16  & 20.24$\times$ \\
\bottomrule
\end{tabular}
\vspace{-12pt}
\end{table}

\section{Conclusion}
In this paper, we propose a learning-based query planning framework for filtered ANN, which dynamically selects between pre-filtering and post-filtering on a per-query basis. With lightweight selectivity estimation and decision models, our method smoothly adapts to diverse datasets and workload, significantly improving query processing performance with minimal overhead. 

\bibliographystyle{ACM-Reference-Format}
\bibliography{citations}

@misc{patel_acorn:_2024,
	title = {{ACORN}: {Performant} and {Predicate}-{Agnostic} {Search} {Over} {Vector} {Embeddings} and {Structured} {Data}},
	shorttitle = {{ACORN}},
	url = {http://arxiv.org/abs/2403.04871},
	doi = {10.48550/arXiv.2403.04871},
	urldate = {2026-01-31},
	publisher = {arXiv},
	author = {Patel, Liana and Kraft, Peter and Guestrin, Carlos and Zaharia, Matei},
	month = mar,
	year = {2024},
	note = {arXiv:2403.04871},
}

@misc{ma_chase:_2025,
	title = {{CHASE}: {A} {Native} {Relational} {Database} for {Hybrid} {Queries} on {Structured} and {Unstructured} {Data}},
	shorttitle = {{CHASE}},
	url = {http://arxiv.org/abs/2501.05006},
	doi = {10.48550/arXiv.2501.05006},
	urldate = {2026-01-31},
	publisher = {arXiv},
	author = {Ma, Rui and Zhang, Kai and He, Zhenying and Jing, Yinan and Wang, X. Sean and Chen, Zhenqiang},
	month = jan,
	year = {2025},
	note = {arXiv:2501.05006},
}

@misc{wang_milvus:_2021,
	address = {Virtual Event China},
	title = {Milvus: {A} {Purpose}-{Built} {Vector} {Data} {Management} {System}},
	isbn = {9781450383431},
	shorttitle = {Milvus},
	url = {https://dl.acm.org/doi/10.1145/3448016.3457550},
	doi = {10.1145/3448016.3457550},
	language = {en},
	urldate = {2026-01-31},
	booktitle = {Proceedings of the 2021 {International} {Conference} on {Management} of {Data}},
	publisher = {ACM},
	author = {Wang, Jianguo and Yi, Xiaomeng and Guo, Rentong and Jin, Hai and Xu, Peng and Li, Shengjun and Wang, Xiangyu and Guo, Xiangzhou and Li, Chengming and Xu, Xiaohai and Yu, Kun and Yuan, Yuxing and Zou, Yinghao and Long, Jiquan and Cai, Yudong and Li, Zhenxiang and Zhang, Zhifeng and Mo, Yihua and Gu, Jun and Jiang, Ruiyi and Wei, Yi and Xie, Charles},
	month = jun,
	year = {2021},
	pages = {2614--2627},
}

@misc{gollapudi_filtered-diskann:_2023,
	address = {Austin TX USA},
	title = {Filtered-{DiskANN}: {Graph} {Algorithms} for {Approximate} {Nearest} {Neighbor} {Search} with {Filters}},
	isbn = {9781450394161},
	shorttitle = {Filtered-{DiskANN}},
	url = {https://dl.acm.org/doi/10.1145/3543507.3583552},
	doi = {10.1145/3543507.3583552},
	language = {en},
	urldate = {2026-01-31},
	booktitle = {Proceedings of the {ACM} {Web} {Conference} 2023},
	publisher = {ACM},
	author = {Gollapudi, Siddharth and Karia, Neel and Sivashankar, Varun and Krishnaswamy, Ravishankar and Begwani, Nikit and Raz, Swapnil and Lin, Yiyong and Zhang, Yin and Mahapatro, Neelam and Srinivasan, Premkumar and Singh, Amit and Simhadri, Harsha Vardhan},
	month = apr,
	year = {2023},
	pages = {3406--3416},
}

@misc{wu_hqann:_2022,
	title = {{HQANN}: {Efficient} and {Robust} {Similarity} {Search} for {Hybrid} {Queries} with {Structured} and {Unstructured} {Constraints}},
	shorttitle = {{HQANN}},
	url = {http://arxiv.org/abs/2207.07940},
	doi = {10.48550/arXiv.2207.07940},
	urldate = {2026-01-31},
	publisher = {arXiv},
	author = {Wu, Wei and He, Junlin and Qiao, Yu and Fu, Guoheng and Liu, Li and Yu, Jin},
	month = jul,
	year = {2022},
	note = {arXiv:2207.07940},
}

@misc{sehgal_navix:_2025,
	title = {{NaviX}: {A} {Native} {Vector} {Index} {Design} for {Graph} {DBMSs} {With} {Robust} {Predicate}-{Agnostic} {Search} {Performance}},
	shorttitle = {{NaviX}},
	url = {http://arxiv.org/abs/2506.23397},
	doi = {10.48550/arXiv.2506.23397},
	urldate = {2026-01-31},
	publisher = {arXiv},
	author = {Sehgal, Gaurav and Salihoglu, Semih},
	month = jun,
	year = {2025},
	note = {arXiv:2506.23397},
}

@misc{li_sieve:_2025,
	title = {{SIEVE}: {Effective} {Filtered} {Vector} {Search} with {Collection} of {Indexes}},
	shorttitle = {{SIEVE}},
	url = {http://arxiv.org/abs/2507.11907},
	doi = {10.48550/arXiv.2507.11907},
	urldate = {2026-01-31},
	publisher = {arXiv},
	author = {Li, Zhaoheng and Huang, Silu and Ding, Wei and Park, Yongjoo and Chen, Jianjun},
	month = jul,
	year = {2025},
	note = {arXiv:2507.11907},
}

@misc{liang_unify:_2025,
	title = {{UNIFY}: {Unified} {Index} for {Range} {Filtered} {Approximate} {Nearest} {Neighbors} {Search}},
	shorttitle = {{UNIFY}},
	url = {http://arxiv.org/abs/2412.02448},
	doi = {10.48550/arXiv.2412.02448},
	urldate = {2026-01-31},
	publisher = {arXiv},
	author = {Liang, Anqi and Zhang, Pengcheng and Yao, Bin and Chen, Zhongpu and Song, Yitong and Cheng, Guangxu},
	month = jun,
	year = {2025},
	note = {arXiv:2412.02448},
}

@misc{wang_wow:_2025,
	title = {{WoW}: {A} {Window}-to-{Window} {Incremental} {Index} for {Range}-{Filtering} {Approximate} {Nearest} {Neighbor} {Search}},
	shorttitle = {{WoW}},
	url = {http://arxiv.org/abs/2508.18617},
	doi = {10.48550/arXiv.2508.18617},
	urldate = {2026-01-31},
	publisher = {arXiv},
	author = {Wang, Ziqi and Zhang, Jingzhe and Hu, Wei},
	month = sep,
	year = {2025},
	note = {arXiv:2508.18617},
}

@misc{bernhardsson_erikbern/ann-benchmarks_2026,
	title = {ann-benchmarks},
	copyright = {MIT},
	url = {https://github.com/erikbern/ann-benchmarks},
	urldate = {2026-02-01},
	author = {Bernhardsson, Erik},
	month = jun,
	year = {2025},
    publisher = {GitHub},
    journal = {GitHub repository},
}

@misc{qdrant_qdrant/ann-filtering-benchmark-datasets_2026,
	title = {ann-filtering-benchmark-datasets},
	url = {https://github.com/qdrant/ann-filtering-benchmark-datasets},
	urldate = {2026-02-01},
	author = {Qdrant},
	month = jun,
	year = {2025},
    publisher = {GitHub},
    journal = {GitHub repository},
}

@misc{qdrant_qdrant/wolt-food-clip-vit-b-32-embeddings_nodate,
	title = {wolt-food-clip-{ViT}-{B}-32-embeddings},
	url = {https://huggingface.co/datasets/Qdrant/wolt-food-clip-ViT-B-32-embeddings},
	urldate = {2026-02-01},
	month = feb,
	year = {2024},
	author = {Qdrant},
    publisher = {Hugging Face},
    journal = {Hugging Facerepository},
}

@misc{li2025attributefilteringapproximatenearest,
      title={Attribute Filtering in Approximate Nearest Neighbor Search: An In-depth Experimental Study}, 
      author={Mocheng Li and Xiao Yan and Baotong Lu and Yue Zhang and James Cheng and Chenhao Ma},
      year={2025},
      eprint={2508.16263},
      archivePrefix={arXiv},
      primaryClass={cs.DB},
      url={https://arxiv.org/abs/2508.16263}, 
}

\end{document}